\documentclass[twocolumn]{autart}    % Enable this line and disable the
\usepackage[T1]{fontenc}

\usepackage[dvips]{epsfig}    % or this line, depending on which
\usepackage{graphics} % for pdf, bitmapped graphics files
\usepackage{epsfig} % for postscript graphics files
\usepackage{times} % assumes new font selection scheme installed
\usepackage{amsmath} % assumes amsmath package installed
\usepackage{amssymb}  % assumes amsmath package installed
\usepackage{cite}
\usepackage{algorithm}
\usepackage{algorithmic}
\usepackage[table]{xcolor}
\usepackage{bbm}
\usepackage{caption}
\usepackage{subcaption}
\usepackage{tikz}
\usepackage{hyperref}
\usepackage{enumerate}
\definecolor{myemphcolor}{cmyk}{0.57,0.55,0,0}

\DeclareMathOperator*{\argmax}{arg\,max}

\newcommand\subscr[2]{#1_{\textup{#2}}}

\definecolor{black}{rgb}{0,0,0}
\definecolor{Red}{rgb}{1,0,0}
\definecolor{Blue}{rgb}{0,0,1}
\definecolor{Green}{rgb}{0,1,0}
\definecolor{magenta}{rgb}{1,0,.6}
\definecolor{lightblue}{rgb}{0,.5,1}
\definecolor{lightpurple}{rgb}{.6,.4,1}
\definecolor{gold}{rgb}{.6,.5,0}
\definecolor{orange}{rgb}{1,0.4,0}
\definecolor{hotpink}{rgb}{1,0,0.5}
\definecolor{newcolor2}{rgb}{.5,.3,.5}
\definecolor{newcolor}{rgb}{0,.3,1}
\definecolor{newcolor3}{rgb}{1,0,.35}
\definecolor{darkgreen1}{rgb}{0, .35, 0}
\definecolor{darkgreen}{rgb}{0, .6, 0}
\definecolor{darkred}{rgb}{.75,0,0}

\xdefinecolor{olive}{cmyk}{0.64,0,0.95,0.4}
\xdefinecolor{purpleish}{cmyk}{0.75,0.75,0,0}

\newcommand{\1}{\mathbf{1}}

\newcommand{\mR}{\mathcal{R}}

\newcommand{\mE}{\mathcal{E}}

\newcommand{\mD}{\mathcal{D}}
\newcommand{\mG}{\mathcal{G}}

\newcommand{\mK}{\mathcal{K}}
\newcommand{\mV}{\mathcal{V}}

\newtheorem{theorem}{\bf Theorem}

\newtheorem{proposition}{\bf Proposition}
\newtheorem{lemma}{\bf Lemma}
\newtheorem{corollary}{\bf Corollary}

\newtheorem{definition}{\bf Definition}
\newtheorem{remark}{\bf Remark}
\newcommand{\mSo}{\subscr{\mathcal{S}}{source}}
\newcommand{\mSoT}{\subscr{\mathcal{S}}{N-source}}

\newcommand{\real}{{\mathbb{R}}}
\newcommand{\realpositive}{{\mathbb{R}}_{>0}}
\newcommand{\integers}{\mathbb{Z}}

\newcommand{\realnonnegative}{{\mathbb{R}}_{\ge 0}}

\newcommand{\oprocendsymbol}{\hbox{$\bullet$}}
\newcommand{\oprocend}{\relax\ifmmode\else\unskip\hfill\fi\oprocendsymbol}

\newcommand{\proofendsymbol}{\hbox{$\Diamond$}}
\newcommand{\proofend}{\relax\ifmmode\else\unskip\hfill\fi\proofendsymbol}

\newcommand{\Rmnum}[1]{\expandafter\@slowromancap\romannumeral #1@}

\newcommand{\indicator}[1]{\mathbbm{1}_{\{ {#1} \} }}

\newcommand{\iset}{$i\in [n]$}{}
\newcommand{\iSet}{$i\in [N]$}

\newcommand{\diag}{\textup{diag}}
\newcommand{\Lie}{\mathcal{L}}

\newcommand{\map}[3]{#1:#2 \rightarrow #3}

\begin{document}

\begin{frontmatter}

\title{Stability of Epidemic Models over Directed Graphs: \\ A Positive Systems Approach\thanksref{footnoteinfo}} % Title, preferably not more
%                                                % than 10 words.
%
\thanks[footnoteinfo]{This paper was not presented at any IFAC meeting. Corresponding author A.~Khanafer Tel. +1-613-501-4234. Research supported in part by an AFOSR MURI Grant FA9550-10-1-0573 and an NSERC Discovery Grant. This manuscript substantially improves and extends the preliminary conference versions \cite{KhanaferBasarGharesifardACC14,KhanaferBasarGharesifardCDC14}.}
\author[UIUC]{Ali Khanafer}\ead{khanafe2@illinois.edu},    % Add the
\author[UIUC]{Tamer Ba\c{s}ar}\ead{basar1@illinois.edu},               % e-mail address
\author[Queens]{Bahman Gharesifard}\ead{bahman@mast.queensu.ca}  % (ead) as shown

\address[UIUC]{Coordinated Science Laboratory, ECE Department, University of Illinois at Urbana-Champaign, USA} % Please supply                                              
\address[Queens]{Department of Mathematics and Statistics, Queen's University, Canada}             % full addresses

                             % keyword list or with the
                                          % help of the Automatica
                                          % keyword wizard

\begin{abstract}                          % Abstract of not more than 200 words.
We study the stability properties of a susceptible-infected-susceptible (SIS) diffusion model, so-called the $n$-intertwined Markov model, over arbitrary directed network topologies. As in the majority of the work on infection spread dynamics, this model exhibits a threshold phenomenon. When the curing rates in the network are high, the disease-free state is the unique equilibrium over the network. Otherwise, an endemic equilibrium state emerges, where some infection remains within the network. Using notions from positive systems theory, {we provide novel proofs for the global asymptotic stability of the equilibrium points in both cases over strongly connected networks based on the value of the basic reproduction number, a fundamental quantity in the study of epidemics.} When the network topology is weakly connected, we provide conditions for the existence, uniqueness, and global asymptotic stability of an endemic state, and we study the stability of the disease-free state. Finally, we demonstrate that the $n$-intertwined Markov model can be viewed as a best-response dynamical system of a concave game among the nodes. This characterization allows us to cast new infection spread dynamics; additionally, we provide a sufficient condition for the global convergence to the disease-free state, which can be checked in a distributed fashion. Several simulations demonstrate our results.
\end{abstract}

\end{frontmatter}

\section{Introduction}
Epidemiological models for disease spread among humans constitute important classes of spread dynamics, as they can potentially provide models for many engineering related phenomena such as the spread of viruses in computer networks \cite{goffman1967communication,kephart1991directed,ganesh2005effect,van2009virus}. There  is a vast literature on various aspects of epidemiological models and the study of infection propagation over networks; we refer the reader particularly to~\cite{pastor2001epidemic,wang2003epidemic,kephart1991directed} and the references therein. Characterization of the stability properties of such diffusion dynamics is a crucial first step towards designing efficient algorithms for controlling their evolutions. Most dynamical epidemiological models, including the $n$-intertwined Markov model \cite{van2009virus,van2013inhomogeneous} studied here, can possess two equilibrium points, under certain conditions: an \emph{disease-free} state at which the network is cured, and an \emph{endemic} state at which the infection persists in the network\cite{lajmanovich1976deterministic,diekmann1990definition,fall2007epidemiological,shuai2013global}. {This has also been observed in time-varying or switching models that allow for abrupt changes in their parameters \cite{rami2014stability}.} A threshold called the basic reproduction number, whose value depends on the curing and infection rates across the network as well as the network topology, determines to which equilibrium point the state of the network will converge \cite{diekmann1990definition}. 

For the $n$-intertwined Markov model, the basic reproduction number, introduced as a critical threshold in~\cite{van2009virus,van2013inhomogeneous}, characterizes this threshold phenomenon. In particular, when the basic reproduction number is less than or equal to $1$, the unique equilibrium is the disease-free state; otherwise, the endemic state emerges. Our aim in this paper is to fully characterize the stability properties of this model over networks with 
directed topologies. {Moreover, we intend to use fundamental results from positive systems theory to construct proofs that could potentially become a starting point for studying the stability of a variety of epidemiological models that share similar characteristics with $n$-intertwined Markov model.}

%Although the equilibria and the critical threshold have been characterized, complete proofs for the asymptotic stability of these equilibrium points have not been provided in the literature. 

\subsection*{Literature review}
A sufficient condition for the stability of the disease-free state over strongly connected digraphs has been established in \cite{preciado2014optimal}. For compartmental susceptible-infected-susceptible (SIS) models, a necessary and sufficient condition for the global asymptotic stability of this equilibrium was presented in \cite{fall2007epidemiological} using a linear Lyapunov function. 
For the same model, the global asymptotic stability of the endemic state over strongly connected directed graphs has been studied in \cite{fall2007epidemiological,ahnglobal,shuai2013global}---see \cite{shuai2013global} for a summary of other approaches to establish this result. The results in \cite{fall2007epidemiological,ahnglobal} rely on the assumption that the state of the model will evolve in the strictly positive quadrant when the state of the network is initialized away from the origin. The result in \cite{shuai2013global} was established using a non-quadratic Lyapunov function, {and by relying on advanced combinatorial results such as Kirchhoff's matrix tree theorem. In contrast, in this paper, using the theory of positive systems, we offer a novel and rigorous proof for the global asymptotic stability of the endemic state over strongly connected digraphs}. This allows us to provide novel results for the stability properties of epidemic dynamics over weakly connected topologies; in all the aforementioned results, the underlying graphs were assumed to be strongly connected (or connected when the graph is undirected). Nonetheless, weakly connected directed graphs are common in practice, and characterizing the equilibrium points as well as their stability properties over these graphs present new challenges in studying epidemiological networks. 

\subsection*{Statement of Contributions}
{The main contributions of this paper are as follows. First, using tools from the theory of positive systems, we characterize the stability properties of the endemic state equilibrium of the $n$-intertwined Markov model over strongly connected digraphs. In particular, we show that when the basic reproduction number is greater than $1$, the endemic state is locally exponentially stable, and when the network is not initialized at the disease-free state, we show that the endemic state is globally asymptotically stable (GAS). Unlike \cite{fall2007epidemiological,ahnglobal}, the proofs we present here do not make any assumption on the evolution of the state, and unlike \cite{shuai2013global}, the stability properties are established using a quadratic Lyapunov function that allows us to avoid relying on advanced combinatorial and graph-theoretic notions.} Using this key construction, our next contribution is to study the existence, uniqueness, and stability properties of the disease-free and endemic states over weakly connected digraphs. 
By studying the input-to-state stability of the network, we provide conditions for a GAS endemic state to emerge over weakly connected digraphs. Unlike endemic states over strongly connected digraphs, we show that at the endemic states emerging over weakly connected graphs a subset of the nodes could be healthy while the rest become infected. 

Finally, we provide a game-theoretic framework that can prescribe more general classes of infection dynamics. Using this model, we show that the $n$-intertwined Markov model prescribes the best-response dynamics of a concave game. This allows us to provide a new condition for the stability of the disease-free state, which can be checked in a distributed way by the nodes.

\subsection*{Organization}
Section~\ref{notation} establishes some mathematical preliminaries required in this paper. In Section~\ref{sec::NIMM}, we recall the $n$-intertwined Markov model, and discuss a connection with a game-theoretic formulation. Sections \ref{sec::directed} and \ref{sec::weakly} contain our results
on the stability of the $n$-intertwined Markov model over, respectively, strongly and weakly connected digraphs. Numerical studies are provided in Section \ref{sec::simulations}. Finally, Section~\ref{sec::conclusion} collects our conclusions and ideas for future work. An Appendix contains technical results that are used in proving some of our main results.

\section{Mathematical Preliminaries} \label{notation}
We start with some terminology and notational conventions. All the matrices and vectors in this paper are real valued. For a set of $ n\in \integers_{\geq 1} $ elements, we use the combinatorial notation $[n]$ to denote $\{1,\hdots,n\}$. The $(i,j)$-th entry of a matrix $X\in \real^{n\times m}$, $n,m\in \integers_{\geq 1}$ is denoted by $x_{ij}$. For two real vectors $x,y \in \real^n$, $ n\in \integers_{\geq 1} $, we write
$x \gg y$ if $x_{i} > y_{i}$ for all \iset, $x\succ y$ if $x_{i}\geq y_{i}$
for all \iset{} but $x\neq y$, and $x\succeq y$ if $x_{i} \geq y_{i}$ for all \iset. We say a vector $x \in \mathbb{R}^n$ is strictly positive if $x \gg 0$. For any vector $x\in \mathbb{R}^{n}$, we define $\subscr{x}{min} := \min_{i\in [n]} x_i$ and $\subscr{x}{max} := \max_{i \in [n]} x_i$. The absolute value of a scalar variable is denoted by $|.|$. We also denote the cardinality of a finite set by $|.|$, and the purpose this operator is being used for will be clear from the context. The set of eigenvalues of a matrix $X$ is denoted by $\sigma(X)$. The spectral radius of a matrix $X \in \mathbb{R}^{n\times n}$ is given by $\rho(X) = \max_{\lambda \in \sigma(X)} |\lambda|$, and its abscissa is given by $\mu(X) = \max_{\lambda \in \sigma(X)} \text{Re}(\lambda)$. When the eigenvalues of a matrix $X$ are real, we denote the largest eigenvalue by $\lambda_1(X)$ and the smallest eigenvalue by $\lambda_n(X)$. The Euclidean norm of a vector is denoted by $\|.\|_2$. The induced $2$-norm of a matrix $X\in \real^{n\times n}$ is given by 
$$
\|X\|_2=\max\limits_{\substack{y\in \real^{n}\\\|y\|_2=1}}\|Xy\|_2=\sqrt{\lambda_1\left(X^TX\right)}.
$$ 

We use the operator $\diag(.)$ for two purposes. When applied to a square matrix $ X\in \mathbb{R}^{n\times n} $, \diag$(X)$ returns a column vector that contains the diagonal entries of $X$. For a vector $x\in \mathbb{R}^n$, $X=\diag(x)$, or $X=\diag(x_1,\hdots,x_n)$, is a diagonal matrix with $X_{ii}=x_i$, \iset. When a diagonal matrix has positive diagonal entries, we call it a positive diagonal matrix. The identity matrix is denoted by $I$, and the all-ones vector is denoted by $\mathbf{1}$. We assume both $I$ and $\1$ have the appropriate dimensions whenever used.

%Let $f: \mathbb{R}^n\to \mathbb{R}^n$ be a continuously differentiable function that defines a dynamical system $\dot{x} = f(x)$, and let $\Eq(f) \subseteq \mathbb{R}^n$ be the set of equilibrium points of this system, i.e., $f(\overline{x}) = 0$ for all $\overline{x} \in \Eq(f)$. The Jacobian matrix of $f$, $J(x) \in \mathbb{R}^{n \times n}$, is given by $J(x) =  \frac{\partial}{\partial x} f(x)$. Assume that the state $x$ of the dynamical system evolves in a compact domain $\mD \subseteq \mathbb{R}^{n\times n}$. We call the continuously differentiable function $V:\mD \to \mathbb{R}$ a Lyapunov function if, for some $\overline{x} \in \Eq(f)$, $V(\overline{x}) =0$ and $V(x)>0$ for all $x \in \mD\setminus \{ \overline{x} \}$. The Lie derivative of $V$ along $f$ is given by
Let $f: \mathbb{R}^n\to \mathbb{R}^n$ be a continuously differentiable function that defines a dynamical system $\dot{x} = f(x)$, and let $\overline{x}$ be an equilibrium point of this system, i.e., $f(\overline{x}) = 0$. The Jacobian matrix of $f$, $J(x) \in \mathbb{R}^{n \times n}$, is given by $J(x) =  \frac{\partial}{\partial x} f(x)$. Let $D\subset \real^{n\times n}$ be a compact domain where the trajectories of the dynamical system $\dot{x} = f(x)$ lie. A continuously differentiable function $V:\mD \to \mathbb{R}$ is a Lyapunov function if, $V(\overline{x}) =0$ and $V(x)>0$ for all $x \in \mD\setminus \{ \overline{x} \}$. The Lie derivative of $V$ along $f$ is given by
\[
\Lie_f V(x) := \frac{d}{dx} V(x)^T f(x).
\]
\subsection*{Matrix Theory}
We call two matrices $X,Y \in \mathbb{R}^{n \times n}$ \emph{similar} if there exists a nonsingular matrix $T \in \mathbb{R}^{n \times n}$ such that $Y = T^{-1}XT$. An important property of similar matrices is that they share the same set of eigenvalues~\cite{horn2012matrix}. Some of our results rely on properties of Metzler and irreducible matrices. A real square matrix $X$ is called Metzler if its off-diagonal entries are nonnegative. We say that a matrix $X \in \mathbb{R}^{n \times n}$ is reducible if there exists a permutation matrix $T$ such that
$$
T^{-1}XT = \left[\begin{array}{cc}Y & Z \\0 & W\end{array}\right],
$$
where $Y$ and $W$ are square matrices, or if $n=1$ and $X=0$ \cite{berman1979nonnegative}. A real square matrix is called irreducible if it is not reducible. A survey on Metzler matrices and their stability properties can be found in \cite{berman1979nonnegative,farina2011positive,bokharaie2012stability}. Hurwitz Metzler matrices have the following equivalent characterizations.
\begin{proposition}[\hspace{-0.1mm}\cite{rantzer2011distributed}] \label{prop::rantzer}
For a Metzler matrix $X\in \mathbb{R}^{n \times n}$, the following statements are equivalent:
\begin{enumerate}[(i)]
\item The matrix $X$ is Hurwitz.
\item There exists a vector $\xi \gg 0$ such that $X\xi \ll 0$.
\item There exists a vector $\nu \gg 0$ such that $\nu^TX \ll 0$.
\item  There exists a positive diagonal matrix $Q$ such that
\begin{equation} \label{eqn::diagStab}
X^TQ+QX= - K,
\end{equation}
where $K$ is a positive definite matrix.
\end{enumerate}
\end{proposition}

The last characterization is often referred to as \emph{diagonal stability}~\cite{berman1979nonnegative,narendra2010hurwitz}.

The Perron-Frobenius (PF) theorem is a fundamental result in spectral graph theory that characterizes some of the properties of the spectra of Metzler and nonnegative matrices, i.e., matrices whose entries are all nonnegative. We first state the PF theorem for irreducible Metzler matrices  \cite[Theorem 17]{farina2011positive}.
%\cite[Theorem 11]{farina2011positive}.
%\begin{theorem}[PF -- Metzler Case] \label{thm::PFM}
%Let $X \in \mathbb{R}^{n\times n}$ be a Metzler matrix. Then
%\begin{enumerate}%[a)]
%\item $\mu(X) \in \sigma(X)$.
%\item If $Xv = \mu(X)v$, then $v\succ 0$.
%\item If $v\gg 0$ is an eigenvector of $X$, then $Xv = \mu(X)v$.
%\end{enumerate}
%\end{theorem}
%\margin{Ali, I don't have this reference yet (I have ordered it). But is there any example provide there that the gap between necessary and sufficient conditions provided by~2 and~3? The next result should be motivated as a case where there is no such gap.}
%When a Metzler matrix is further irreducible, the following result applies
\begin{theorem}[PF -- Irreducible Metzler Case] \label{thm::PFMI}
Let $X \in \mathbb{R}^{n\times n}$ be an irreducible Metzler matrix. Then
\begin{enumerate}[(i)]
\item $\mu(X)$ is an algebraically simple eigenvalue of $X$.
\item Let $v_F$ be such that $Xv_F = \mu(X)v_F$. Then $v_F$ is unique (up to scalar multiple) and $v_F \gg 0$.
\item If $v\succ 0$ is an eigenvector of $X$, then $Xv = \mu(X)v$, and, hence, $v$ is a scalar multiple of $v_F$.
\end{enumerate}
\end{theorem}

For irreducible nonnegative matrices, the following version of the PF theorem applies \cite[Theorem 8.2.11]{horn2012matrix}.
\begin{theorem}[PF -- Irreducible Nonnegative Case] \label{thm::PFNI}
Let $X \in \mathbb{R}^{n\times n}$ be an irreducible nonnegative matrix. Then
\begin{enumerate}[(i)]
\item $\rho(X)>0$.
\item $\rho(X)$ is an algebraically simple eigenvalue of $X$.
\item If $Xv = \rho(X)v$, then $v\gg 0$.
\end{enumerate}
\end{theorem}

\subsection*{Graph Theory}
A \emph{directed graph}, or \emph{digraph}, is a pair $\mG = (\mV,\mE)$, where $\mV$ is the set of nodes
and $ \mE \subseteq \mV\times \mV $ is the set of edges. Given $\mG$, we denote an edge from
node $i \in \mV$ to node $j\in \mV$ by $(i,j)$. We say node $i \in \mV$ is a neighbor of node $j \in \mV$ if and only if $(i,j) \in \mE$. When $(i,j)\in \mE$ if and only if $(j,i)\in \mE$, we call the graph \emph{undirected}. For a graph with $n \in \integers_{\geq 1}$ nodes, we associate an adjacency matrix $A \in \mathbb{R}^{n \times n}$ with entries $a_{ij}\in \realnonnegative$, where $a_{ij} =0$ if and only if $(i,j) \notin \mE$. For undirected graphs, the adjacency matrix is symmetric, i.e., $A = A^T$. 

In a digraph, a directed path is a collection of nodes $\{i_1,\hdots, i_\ell \} \subseteq \mV$, $\ell \in  \integers_{> 1} $, such that $(i_k,i_{k+1}) \in \mE$ for all $k \in [\ell-1]$. A digraph is \emph{strongly connected} if there exists a directed path between any two nodes in $\mV$. A strongly connected component (SCC) of a graph is a subgraph which itself is strongly connected. A path in an undirected graph is defined in a similar manner. We call an undirected graph \emph{connected} if it contains a path between any two nodes in $\mV$. A digraph is called \emph{weakly connected} if when every edge in $\mE$ is viewed as an undirected edge, the resulting graph is a connected undirected graph. We call a graph, whether it is directed or undirected, \emph{disconnected} if it contains at least two isolated subgraphs. Throughout this paper, when the graph $\mG$ is directed, we assume that it is either strongly or weakly connected. When $\mG$ is undirected, we assume that it is connected.

A directed acyclic graph (DAG) is a digraph with no directed cycles. A node $i\in \mV$ is called a source node if $\sum_{j\neq i} \indicator{a_{ji} \neq 0} = 0$, and it is called a sink node if $\sum_{j\neq i} \indicator{a_{ij} \neq 0} = 0$, where $\indicator{a_{ij} \neq 0} = 1$ if and only if $a_{ij} \neq 0$, and is zero otherwise. A DAG can have multiple sources and multiple sinks. For a given graph $\mG$, let $\mSo$ denote the set of source nodes, and let $\mSoT$ be the set of all nodes $i$ in $\mG$ such that $a_{ji} \neq 0$ for some $j \in \mSo$.

\section{The $n$-Intertwined Markov Model} \label{sec::NIMM}

In this section, we recall the heterogeneous $n$-intertwined Markov model
that has recently been proposed~\cite{van2009virus,van2013inhomogeneous}. This model is related to the so-called
multi-group SIS model that was proposed earlier in \cite{lajmanovich1976deterministic};
see also~\cite{fall2007epidemiological,shuai2013global}. We prescribe the infection model over a directed graph
$\mG = (\mV,\mE)$ with $n$ nodes, where $\mV$ is the set of nodes, and $\mE$ is the set of edges. Each node in the network has two states: infected or cured. The curing and infection of a given node $i \in \mV$ are described by two independent Poisson processes with rates $\delta_i$ and $\beta_i$, respectively. Throughout the paper, we assume that $\delta_i>0$ and $\beta_i>0$. The transition rates between the
 healthy and infected states of a given node's Markov chain depend on its curing rate as well as  
 the infection probabilities among its neighbors. A mean-field approximation is introduced to
 ``average" the effect of infection probabilities of the neighbors on the infection probability of
 a given node. This approximation yields a dynamical system that describes the
 evolution of the probability of infection of node $i\in \mV$ and is central to our upcoming developments. We briefly review this dynamical system next.
 
Let $p_i(t) \in [0,1]$ be the infection probability of node $i \in \mV$ at time $ t\in \realnonnegative$, and let $p(t) = [p_1(t),\hdots,p_n(t)]^T$. Also, let $D = \diag(\delta_1,\hdots,\delta_n)$, $P(t) = \diag(p(t))$, and $B = \diag(\beta_1,\hdots,\beta_n)$. The $n$-intertwined Markov model is prescribed by the mapping $ \map{\Phi}{\real^n}{\real^n} $, where
\begin{eqnarray} \label{eqn::SISmodelMatrix}
\dot{p}(t) & = & \Phi(p(t))  \nonumber \\
& := & (A^TB - D)p(t)-P(t)A^TBp(t).
\end{eqnarray}
It can be shown that when $p(0)\in[0,1]^n$, $p(t)\in [0,1]^n$, for all $t\in \mathbb{R}_{> 0}$ \cite{van2009virus}. Hereinafter, for most parts, we will drop the time index for notational simplicity.

\subsection{Equilibrium States of the $n$-Intertwined Markov Model}

We next focus on characterizing the set of equilibria of the dynamical system~\eqref{eqn::SISmodelMatrix}. We give this characterization using the so-called \emph{basic reproduction number}, denoted by $\mR_0$, which is defined as the expected number of infected nodes produced in a completely susceptible population due to the infection of a neighboring node~\cite{diekmann1990definition}. For the $n$-intertwined Markov model, the basic reproduction number was found in~\cite{van2013inhomogeneous}, where it was called the ``critical threshold", to be equal to
$$
\mR_0 = \rho(D^{-1}A^TB).
$$
For connected undirected graphs, it is shown in \cite{van2013inhomogeneous} that the disease-free state is the unique equilibrium for the $n$-intertwined Markov model when $\mR_0 \leq 1$. When $\mR_0>1$, in addition to the disease-free equilibrium, an endemic equilibrium, denoted by $p^\star$, emerges. In fact, it is shown that $p^\star \gg 0$. We call a strictly positive endemic state \emph{strong}. When $p^\star \succ 0$, we call it a \emph{weak} endemic state. 
A recursive expression for the endemic state $p^\star$ is provided in \cite{van2013inhomogeneous}, which is shown to depend on the problem parameters only: $A$, $\delta_i$, $\beta_i$, $i \in \mV$. To arrive at this expression, consider the steady-state equation
\begin{equation} \label{eqn::steadystateEqn}
0 = (A^TB - D)p - PA^TBp.
\end{equation}
Define $\xi_i := \sum_{j\neq i}a_{ji}\beta_jp_j$ and $\xi_i^\star := \sum_{j\neq i}a_{ji}\beta_jp_j^\star$, $i \in \mV$. We can then write $p_i^\star$ as
\begin{equation} \label{eqn::pStar}
p_i^{\star} = \frac{\xi_i^\star}{\delta_i + \xi_i^\star} = 1 - \frac{\delta_i}{\delta_i + \xi_i^\star}, \quad i \in \mV.
\end{equation}
Since we assumed that $\delta_i>0$, we conclude that $p_i^\star < 1$, for all $i \in \mV$. We can then re-write (\ref{eqn::steadystateEqn}), evaluated at $p^\star$, in the following form:
\begin{equation} \label{eqn::sspStarDirected}
A^TBp^\star = (I-P^\star)^{-1}Dp^\star,
\end{equation}
where $P^\star = \diag(p^\star)$.

\subsection{The $n$-Intertwined Markov Model as a Concave Game} \label{subsec::game}

In this subsection, we demonstrate that the $n$-intertwined Markov model 
can be cast as the best response dynamical system associated with a noncooperative game. 
An important by-product of this study is the development of a larger class of infection dynamics with reasonable convergence properties. Further, our exposition provides a decision-based interpretation to virus spread models, which are often based on the theory of Markov chains. Although our focus here is the study of virus spread, our model can be applied to other diffusion phenomena such as the spread of spam in computer networks.

To this end, consider a digraph $\mG=(\mV,\mE)$ with $n$ nodes, and let $0 \leq x_i \leq 1$ be the rate with which node $i$ sends messages. We associate an objective function, denoted by $f_i: \mathbb{R}^n \to \mathbb{R}$, to node $ i $ that is comprised of a local utility function $U_i: [0,1] \to \mathbb{R}$, and a component that encapsulates the influence of the neighboring nodes. The influence of node $j$ on node $i$ is described via the function $\tilde{g}_{ji}: [0,1]\times[0,1]\to \mathbb{R}$, where $\tilde{g}_{ji} \equiv 0$ if and only if $(j,i) \notin \mE$. We can then write the objective function of node $i$ as
\begin{equation} \label{eqn::utilityGen}
f_i(x_i,x_{-i}) = U_i(x_i) + \sum_{j \neq i} \tilde{g}_{ji}(x_i,x_j).
\end{equation}

Each node is interested in maximizing its own objective function $f_i$. Formally, we can write the problem of the $i$-th agent as
\begin{equation}
\max_{0 \leq x_i \leq 1} f_i(x_i, x_{-i}), \quad \text{for each fixed } x_{-i}. \label{concaveGame}
\end{equation}
When $f_i$ is concave in $x_i$, and because the objective function of each player depends also on the actions of other players, problem (\ref{concaveGame}) describes a concave game \cite{rosen1965existence,BasarOlsder}.

The solution concept we are interested in studying here is the pure-strategy Nash equilibrium (PSNE). %~\bah{[Citation to Rosen Here]}
\begin{definition}[\hspace{-0.05mm}\cite{BasarOlsder}]
The vector $x^\star \in [0,1]^n$ constitutes a PSNE if, for all $ i \in \mV$, the inequality
\[
f_i(x_i^\star,x^\star_{-i}) \geq f_i(x_i,x^\star_{-i})
\]
is satisfied for all $x_i \in [0,1]$.
\end{definition}
Note that under the PSNE, no agent has any incentive to unilaterally deviate from the solution $x^\star$. The next proposition establishes the existence and uniqueness of the PSNE for the game in (\ref{concaveGame}), when the game is concave.
\begin{proposition}[\hspace{-0.05mm}\cite{rosen1965existence}] \label{prop::existUnique}
For  each $i\in \mV$, let $f_i(x_i, x_{-i})$ in \eqref{eqn::utilityGen} be strictly concave in $x_i\in [0,1]$, for every $x_j\in [0,1], j\in \mV, j\neq i$. Then the resulting concave game in \eqref{concaveGame} admits a unique PSNE under the following diagonal dominance condition:
\begin{eqnarray} 
2\left| \frac{\partial^2}{\partial x_i^2} U_i(x_i)\right| & > & \sum_{j\neq i} \left| \frac{\partial}{\partial x_j}\frac{\partial}{\partial x_i} \tilde{g}_{ij}(x_i,x_j) \nonumber \right. \\
&&\left. +\frac{\partial}{\partial x_j}\frac{\partial}{\partial x_i}  \tilde{g}_{ji}(x_j,x_i) \right|. \label{eqn::diagDomCondn}
\end{eqnarray}
\end{proposition}
%We note that an interesting form of the influence function $\tilde{g}_{ji}$ is the following:
%\begin{equation} \label{eqn::utility}
%\tilde{g}_{ji}(x_i,x_j) = a_{ji}x_ig_{ji}(x_j).
%\end{equation}
%The benefit of working with the particular structure in (\ref{eqn::utility}) is twofold: (i) since $x_i$ scales the total influence of the neighboring nodes, $\sum_{j \neq i} a_{ji}g_{ji}(x_j)$, this form highlights the fact that $x_i$ is a rate; (ii) the second derivative of $f_i$ with respect to $x_i$ is independent of $x_{-i}$, which allows us to design concave games by selecting $U_i$ to be concave in $x_i$.
%\begin{pf}
%This is essentially Rosen's result. We only need to argue that the result still holds when the projection operator is used.
%\end{pf}

The following lemma establishes a relationship between virus spread in networks and concave games. In the virus spread case, the probability of infection $p_i$ plays the role of the transmission rate $x_i$.
\begin{lemma} \label{lem::SISasGame}
The dynamics of the $n$-intertwined Markov model are best-response dynamics of a concave game
among the nodes, where the decision variable of node $i \in \mV$ is $p_i\in [0,1]$, and its objective function is given by
\begin{equation} \label{eqn::SISutility}
f_i(p_i,p_{-i}) = -\frac{\delta_i}{2}p_i^2 + p_i\left(1-\frac{p_i}{2}\right) \sum_{j\neq i}a_{ji}\beta_j p_j.
\end{equation}
\end{lemma}
\begin{pf}
Recall the objective functions defined in (\ref{eqn::utilityGen}). Let $U_i(p_i) = -\frac{\delta_i}{2}p_i^2$ and $\tilde{g}_{ji}(p_i,p_j) = p_i(1-\frac{p_i}{2}) a_{ji}\beta_j p_j$, $i \in \mV$. We then obtain
\[
\frac{\partial^2}{\partial p_i^2} f_i(p_i,p_{-i}) = -\delta_i -  \sum_{j\neq i} a_{ji}\beta_j p_j < 0, \quad i \in \mV,
\]
which shows that the $f_i$'s are strictly concave in self variables. It is now not hard to see that the dynamics of the $n$-intertwined Markov model (\ref{eqn::SISmodelMatrix}) correspond to the gradient flow dynamics when the agents aim at maximizing their own objective functions (\ref{eqn::SISutility}). \qed
\end{pf}
\section{Stability of Epidemic Dynamics over Strongly Connected Graphs} \label{sec::directed}
We start by studying the stability properties of the $n$-intertwined 
model over directed graphs with strongly connected topologies.

\subsection{Stability of the Disease-Free State} \label{sec::allhealthy}

As a stepping stone, we first provide an alternative proof for the necessary and sufficient condition for the global asymptotic stability of the disease-free state, see~\cite{fall2007epidemiological,preciado2014optimal}, using the theory of positive systems. As we will see shortly, the proof strategy provided here is essential in some of our upcoming 
results. 

\begin{proposition} \label{prop::unstableDirected}
Suppose $ \mathcal{G} = (\mV,\mE)$ is a strongly connected digraph. The disease-free equilibrium is GAS if and only if $\mR_0 \leq 1$.
\end{proposition}
\begin{pf}
Note that the matrix $A^TB-D$ is Metzler, because the entries of $A^TB$ are nonnegative. Using the convergent regular splitting property of Metzler matrices, it can be shown that $\mR_0 < 1$ if and only if $\mu(A^TB-D)< 0$, and $\mR_0 = 1$ if and only if $\mu(A^TB-D)= 0$~\cite[Theorem 2.3]{berman1979nonnegative}.

As a result, when $\mR_0 < 1$, the matrix $A^TB-D$ is Hurwitz. Since it is also Metzler, by Proposition \ref{prop::rantzer}(iv), there exists a positive diagonal matrix $R_1$ satisfying $(A^TB-D)^TR_1+R_1(A^TB-D) = -K$, where $K$ is a positive definite matrix. Consider the Lyapunov function $V_1(p) = p^TR_1p$. Using~\eqref{eqn::SISmodelMatrix}, we have
\begin{eqnarray}
\Lie_\Phi V_1(p) & = & p^T((A^TB-D)^TR_1+R_1(A^TB-D))p \nonumber \\
& -& 2p^TR_1PA^TBp \nonumber \\
&\leq&  p^T((A^TB-D)^TR_1+R_1(A^TB-D))p \nonumber \\
& = &-p^TKp \leq \lambda_1(-K) \|p\|_2^2 < 0, \quad p \neq 0,\label{eqn::expStab}
\end{eqnarray}
where the first inequality follows because $p^TR_1PA^TBp \geq 0$, for all $p \in [0,1]^n$, and (\ref{eqn::expStab}) follows because $K$ is positive definite. This implies that the disease-free state is GAS.
 
When $\mR_0 = 1$, we have $\mu(A^TB-D)=0$. Since $\mG$ is strongly connected, it follows that $A^TB-D$ is irreducible \cite{berman1979nonnegative}. Recalling that $A^TB-D$ is also Metzler, we conclude from Lemma \ref{lem::semi} that there exists a positive diagonal matrix $R_2$ such that $(A^TB-D)^TR_2+R_2(A^TB-D)$ is negative semidefinite. Using the Lyapunov function $V_2(p)=p^TR_2p$, we can write
\begin{eqnarray}
\Lie_\Phi V_2(p) & = & p^T((A^TB-D)^TR_2+R_2(A^TB-D))p \nonumber \\
& -& 2p^TR_2PA^TBp \nonumber \\
&\leq& -2p^TR_2PA^TBp.  \nonumber
\end{eqnarray}
We next prove that $p^TR_2PA^TBp = 0$ if and only if $p = 0$. Since $R_2$ is a positive diagonal matrix, we have that $p^TR_2PA^TBp = 0$ if and only if 
\begin{equation}\label{eq:aux1-proof}
p_i^2 \sum_{j \neq i} a_{ji}\beta_jp_j = 0,
\end{equation}
for all $i \in \mV$. Assume that there is a solution $p$ that satisfies $p^TR_2PA^TBp = 0$ at some time $t_0 \in \realnonnegative$, and let $p_i(t_0) \neq 0$ for some $i \in \mV$. Then, by continuity of the state $p$, there exists an interval $\tau = [t_0,t_0+\delta]$, $\delta > 0$, such that $p_i(t) \neq 0$, for all $t \in \tau$. Using~\eqref{eq:aux1-proof}, we hence conclude that for all $j\in \mV$ that are neighbors of $ i $, i.e., $a_{ji}\neq 0$, we must have that $p_j(t) = 0$ and $\dot{p}_j(t) = 0 $ for all $t \in \tau$, for all $j\in \mV$ with $a_{ji}\neq 0$. Then, for some $j \in \mV$ such that $a_{ji} \neq 0$, we have $\dot{p}_{j}(t) = \sum_{k\neq j} a_{kj}\beta_k p_k(t) = 0$, for all $t\in \tau$. This implies that $p_k(t) = 0$ for all $t\in \tau$ and for all $k\in \mV$ such that $a_{kj}\neq 0$. By repeating this argument, we conclude that $p_{l}(t) = 0$ for all $t\in \tau$ for any node $l\in \mV$ from which there is a directed path to node $j$. Since $\mG$ is strongly connected, there is a directed path from node $i$ to node $j$, and we must then have $p_i(t) = 0$ for all $t\in \tau$, which contradicts our initial hypothesis. It then follows that $p^TR_2PA^TBp = 0$ if and only if $p \equiv 0$. Hence, the disease-free state is GAS. This proves the sufficiency part.

We will show necessity by proving the contrapositive. The Jacobian matrix of the vector field in \eqref{eqn::SISmodelMatrix} evaluated at the origin is given by $J(0) = A^TB-D$. If $\mR_0>1$, we have $\mu(A^TB-D) > 0$, and we conclude by Lyapunov's indirect method that the original nonlinear system is not stable. This proves that $\mR_0 \leq 1$ is also necessary for the disease-free equilibrium to be asymptotically stable. \qed
\end{pf}

It is worth noting that, when $\mR_0<1$, the proof of the global asymptotic stability of the disease-free state does not rely on the strong connectivity assumption. This is also true for the instability proof, when $\mR_0 > 1$. We only used the strong connectivity of the graph to prove global asymptotic stability when $\mR_0 = 1$. {Further, note that $\mR_0$ provides a sharp threshold for the stability of the disease-free equilibrium. To characterize the speed of convergence, one should provide an upper bound for $\mu(A^TB-D)$; see \cite[Proposition 1]{preciado2014optimal}}. 

%\begin{remark} \longthmtitle{Graph connectivity and the stability of the disease-free state} \label{rem::0}

%\oprocend
%\end{remark}
%\begin{remark}\longthmtitle{Comparison with existing results}
%In~\cite{fall2007epidemiological}, and using a linear Lyapunov function, a proof is given for the stability of the healthy state in the so-called compartmental SIS models. It is not hard to observe that the dynamical system~\eqref{eqn::SISmodelMatrix} can be cast as a compartmental SIS model, and, hence, Proposition~\ref{prop::unstableDirected} and the result in \cite{fall2007epidemiological} are equivalent. It is worth mentioning that the sufficiency of $\mu(A^TB-D) < 1$ for the global asymptotic stability of the disease-free state is also proved in \cite{preciado2014optimal}, using the comparison lemma.
%
%As we will shortly see, the proof strategy provided above is essential in some of the upcoming results in this paper. Additionally, when $R_0 < 1$, the proof above relies on a quadratic Lyapunov function, which enables us to determine the convergence rate. In particular, we the Lie derivative of the Lyapunov function we used is upper bounded by a quadratic function, which implies that the convergence rate is exponential.
%\oprocend
%\end{remark}

\subsection{Existence and Stability of an Endemic State} \label{sec::endemic}
%\subsubsection{Strongly Connected Digraphs}
In this section, we use notions from positive systems theory to prove the local and global asymptotic stability of an endemic state over strongly connected digraphs. We first note that the existence of a unique 
endemic state for~\eqref{eqn::SISmodelMatrix} over strongly connected digraphs can be concluded from~\cite[Section 2.2]{fall2007epidemiological}, as stated next.
%for compartmental SIS model, 

\begin{proposition}[{\hspace{-0.05mm}\cite{fall2007epidemiological}}]\label{prop::pStarUniqueDirected}
Let $\mG = (\mV,\mE)$ be a strongly connected digraph. Then, a unique strong endemic state $p^\star \gg 0$ exists if and only if $\mR_0 > 1$.
\end{proposition}

Next, we compute the Jacobian of $\Phi$, given by~\eqref{eqn::SISmodelMatrix}, at $p^\star$. Note that
\begin{eqnarray*} 
J_{ii}(p^\star)  & = & \frac{\partial}{\partial p_i} \Phi_i(p^\star)= -(\delta_i + \xi_i^\star), \quad i \in \mV,\\
 J_{ij}(p^\star) & = & \frac{\partial}{\partial p_j} \Phi_i(p^\star)= (1-p_i^\star)a_{ji}\beta_{j}, \quad j\neq i, j \in \mV,
\end{eqnarray*}
where $\Phi_i(p^\star)$ is $i$-th entry of $f(p^\star)$. Using the definition of $p^\star$ in (\ref{eqn::pStar}), we realize that $J_{ii}(p^\star) = -\delta_i/(1-p_i^\star)$, $i \in \mV$. As a result, we conclude that
\begin{equation} \label{eqn::jacobPstarDirected}
J(p^\star) = -(I-P^\star)^{-1}D+(I-P^\star)A^TB.
\end{equation}
Our first result establishes the local stability of $p^\star$.

\begin{theorem}\label{thm:lesDirected}%\longthmtitle{Local exponential stability of the strong endemic state in strongly connected digraphs}
Suppose that $\mG = (\mV,\mE)$ is a strongly connected digraph and that $\mR_0 > 1$.
Then, the strong endemic state $p^\star$ is locally exponentially stable.
%In a strongly connected digraph $\mG$, the endemic state $p^\star$ is locally exponentially stable when $\lambda_1(A^TB-D)>0$.
\end{theorem}
\begin{pf}
We invoke Lyapunov's indirect method. Since $ \mG $ is strongly connected, $A$ is irreducible. From (\ref{eqn::sspStarDirected}), we deduce that $Dp^\star = (I-P^\star)A^TBp^\star$. We can then write
\begin{eqnarray*}
J(p^\star)p^\star & = &  -A^TBp^\star+(I-P^\star)A^TBp^\star\\
&=& -P^\star A^TBp^\star \ll 0,
\end{eqnarray*}
where the last strict inequality follows because $p^\star \gg 0$, $B$ is a positive diagonal matrix, and $A$ is irreducible.
The matrix $J(p^\star)$ is Metzler, because its off-diagonal entries are nonnegative. Then, using Proposition \ref{prop::rantzer}(ii), we conclude that $J(p^\star)$ is Hurwitz. \qed
\end{pf}

We are now in a position to state the following result.

\begin{theorem}\label{thm::pStarGASDirected}%\longthmtitle{global asymptotic stability of the strong endemic state in strongly connected digraphs}  
Let $\mG=(\mV,\mE)$ be a strongly connected digraph, and assume that $p(0) \neq 0$. If $\mR_0>1$, then
the strong endemic state $p^\star$ is GAS.
\end{theorem}
\begin{pf}
Recall that $p(t) \in [0,1]^n$ for all $t\in \realnonnegative$. When $\mR_0>1$, Proposition~\ref{prop::unstableDirected}
implies that the disease-free equilibrium is unstable. Therefore, under this condition, the set $ W=[0,1]^n\backslash \{0\} $ is invariant under the evolutions
of~\eqref{eqn::SISmodelMatrix}. %Let $ p(0)\in W $ be any initial condition.

Next, define the state $\tilde{p} = p - p^\star$. Let $\tilde{P} = \diag(\tilde{p})$. The dynamics of $\tilde{p}$ can then be written as follows:
\begin{eqnarray*}
\dot{\tilde{p}} & = & (A^TB-D)(\tilde{p}+p^\star) -(\tilde{P}+P^\star)A^TB(\tilde{p}+p^\star) \nonumber \\
& = &(-D+(I-P^\star)A^TB)\tilde{p}-\tilde{P}A^TBp. %\label{eqn::ptilde}
\end{eqnarray*}
Define the matrix $\Lambda(p^\star) :=   -D + (I-P^\star)A^TB$, and note that the off-diagonal entries of $\Lambda(p^\star)$ are nonnegative; hence, $\Lambda(p^\star)$ is a Metzler matrix. Since $\mG$ is strongly connected, the matrix $\Lambda(p^\star)$ is also irreducible. From (\ref{eqn::sspStarDirected}), it follows that $\Lambda(p^\star) p^\star = 0$, and since $p^\star$ is strictly positive, it follows from Theorem \ref{thm::PFMI} that $\mu(\Lambda(p^\star)) = 0$. Thus, it follows from Lemma \ref{lem::semi} that there exists a positive diagonal matrix $R$ such that the matrix $\Lambda(p^\star)^TR+ R\Lambda(p^\star)$ is negative semidefinite.

Consider the Lyapunov function $V(\tilde{p})= \tilde{p}^TR\tilde{p}$. We have
\begin{eqnarray*}
\Lie_{\Phi} V(\tilde{p}) & = & \tilde{p}^T(\Lambda(p^\star)^TR + R\Lambda(p^\star))\tilde{p} - 2\tilde{p}^T\tilde{P}RA^TBp \\
& \leq & - 2\tilde{p}^TR\tilde{P}A^TBp = - 2\tilde{p}^T\tilde{P}RA^TBp,
\end{eqnarray*}
where the inequality follows because $\Lambda(p^\star)^TR + R\Lambda(p^\star)$ is negative semidefinite, and the last equality follows because $\tilde{P}$ and $R$ commute, since they are both diagonal matrices.

We next prove that $p^TRPA^TBp = 0$ if and only if $p = p^\star$. Since $R$ is a positive diagonal matrix, we have $\tilde{p}^T\tilde{P}RA^TBp = 0$ if and only if $\tilde{p}_i^2 \sum_{j \neq i} a_{ji}\beta_j p_j=0$, for all $i\in \mV$. Assume that there is a vector $p$ that satisfies $\tilde{p}^T\tilde{P}RA^TBp = 0$ while $p_i \neq p_i^\star$, for some $i\in \mV$. We then must have $\sum_{j \neq i} a_{ji}\beta_j p_j=0$, which implies that $p_j = 0$ for all $j \in \mV$ such that $a_{ji} \neq 0$. Then, for some $j \in \mV$ for which $a_{ji} \neq 0$, we must also have $\sum_{k \neq j} a_{kj}\beta_k p_k=0$, because $p_j=0 < p_j^\star$. By repeating this argument, we conclude that $p_l = 0$ for any node $l \in \mV$ from which there is a directed path to node $j$. Since $\mG$ is strongly connected, there is a directed path from node $i$ to node $j$, and we must have $p_i = 0$. This implies that $p=0$, which contradicts our initial assumption. Therefore, since the set $ W $ is invariant under~\eqref{eqn::SISmodelMatrix}, we have that $  \dot{V}(\tilde{p}) = 0  $ if and only if $ p=p^\star $. \qed
\end{pf}

\begin{remark}
{The novelty in our proof lies at the utilization of notions from positive systems theory, which enables us to  avoid the need to make combinatorial arguments about the underlying graph structure as in the proof that utilizes a logarithmic Lyapunov function in~\cite{shuai2013global}. Similar to the proof in~\cite{shuai2013global}, our proof enjoys the advantage of relying on a single Lyapunov function as opposed to the proof in~\cite{lajmanovich1976deterministic} that constructs two Lyapunov functions to prove this result.}

A proof for a weaker statement is established in~\cite{fall2007epidemiological,ahnglobal}, where
it is assumed that for $p(0) \neq 0$, there exists a time $ T \in \realpositive$ such that $p(t) \in (0,1]^n$ for all $t \geq T$.

In addition to the useful characteristics of using a quadratic Lyapunov function for studying additional properties such as convergence rates, our proof allows for establishing the stability properties of the equilibrium points over weakly connected digraphs in the next section.
\oprocend
\end{remark}

\subsection{A Simplified Stability Condition through a Game-Theoretic Perspective}
The game-theoretic connection we established in Lemma \ref{lem::SISasGame} enables us to provide a simplified condition for the global asymptotic stability of the disease-free state. In particular, by applying the diagonal dominance condition in (\ref{eqn::diagDomCondn}) to (\ref{eqn::SISutility}), we obtain the following sufficient condition:
\begin{equation} \label{eqn::SISmodelCnvrgCondnRosen}
\frac{1}{2} \sum_{j \neq i}a_{ij}\beta_j < \delta_i, \quad \text{for all } i \in \mV.
\end{equation}

Recall that the conditions $\mR_0<1$ and $\mu(A^TB-D)<0$ are equivalent. Note the similarities between the conditions $\mu(A^TB-D)<0$ and (\ref{eqn::SISmodelCnvrgCondnRosen}). The two conditions are related by the Gershgorin Circle Theorem. While (\ref{eqn::SISmodelCnvrgCondnRosen}) is more restrictive than $\mu(A^TB-D)<0$, it is linear and easier to compute. More importantly, condition (\ref{eqn::SISmodelCnvrgCondnRosen}) can be checked in a distributed fashion, which makes it more suitable for the design of distributed algorithms. %Hence, we have been able to provide a sufficient ``distributed" condition for the stability of the disease-free state of the $n$-intertwined Markov model using the uniqueness condition of the PSNE of the underlying concave game.

\section{Stability of Epidemic Dynamics over Weakly Connected Graphs} \label{sec::weakly}

In this section, we study the stability properties of the $n$-intertwined Markov model over weakly
connected graphs. This class is of great importance, since it is conceivable that in many practical scenarios there exist connected components that collectively serve as an infection source, but are
not affected by the rest of the nodes. Such scenarios cannot be captured by strongly connected topologies.

%Studying epidemiological models over
%weakly connected digraphs can yield new epidemic behaviors that do not
%emerge over strongly connected digraphs. For example, as we will show by providing an example, the general belief that
%the endemic state is always strictly positive does not hold for weakly connected graphs,
%and weak endemic states can emerge.

We start by introducing some notations. When the graph $\mG$ is weakly connected, its adjacency matrix can be transformed
into an upper triangular form using an appropriate labeling of the nodes.
%\margin{Ali, I would use the word vertices/vertex all over the place instead of nodes}
%\margin{define SCC somewhere in the prelim}
%permutation of the nodes.
Assuming that $\mG=(\mV,\mE)$ contains $N \in \integers_{\geq 1}$ strongly connected components,
we can write
\[
A = \left[\begin{array}{cccc}A_{11} & A_{12} & \hdots & A_{1N} \\0 & A_{22} & A_{23} & \hdots \\ \vdots & \ddots & \ddots & \ddots \\ 0 & \hdots & 0 & A_{NN}\end{array}\right],
\]
where $A_{ii}$ are irreducible for all \iSet, and, hence, correspond to SCCs in $\mG$~\cite{berman1979nonnegative}. For notational simplicity, we will use $A_i$ instead of $A_{ii}$. The matrices $A_{ij}$, $j \neq i$ are not necessarily irreducible. We denote an SCC of $\mG$ by $\mG_i = (\mV_i, \mE_i )$, \iSet, where $\cup_{i=1}^N \mV_i = \mV$ and $\cup_{i=1}^N \mE_i = \mE$. For each \iSet, we introduce the positive diagonal matrices $D_i$, $B_i$ which contain, respectively, the curing and infection rates of the nodes in $\mV_i$ along their diagonals. We introduce the partial order '$\prec$' among SCCs, and we write $\mG_i \prec \mG_j$, for some $i,j \in [N]$, if there is a directed path from $\mG_i$ to $\mG_j$ but not vice versa.

For a given \iSet, we denote the state of the nodes in $\mG_i$ by $q_i \in \mathbb{R}^{|\mV_i|}$ and the state of the $k$-th node in $\mV_i$ by $q_{i,k} \in \mathbb{R}$. The state, $p$, of the entire network is given by $p=[q_1^T,\hdots,q_N^T]$. Let $c_{i} = \sum_{j\neq i} A_{ji}^TB_jq_j \in \mathbb{R}^{|\mV_i|}$, \iSet, be the input infection from the nodes in $\mG \backslash \mG_i$. We can now write the dynamics of the nodes in $\mG_i$, \iSet, given by the mapping $ \map{\tilde{\Phi}_i}{\real^{|\mV_i|} \times \real^{|\mV_i|}}{\real^{|\mV_i|}} $, as
\begin{eqnarray} \label{eqn::dqGi}
\dot{q}_i & = & \tilde{\Phi}_i(q_i,c_i) \nonumber \\
& := & (A_i^TB_i-D_i)q_i - Q_iA_i^TB_iq_i + (I-Q_i) c_{i},
\end{eqnarray}
where $Q_i = \diag(q_i)$. When an SCC comprises a single node, $A_i^TB_i-D_i$ is equal to $-\delta_i$. In what follows, we say $\mG_i$ is stable to mean that the dynamics (\ref{eqn::dqGi}) are stable. When an endemic state $p^\star$ emerges over the graph $\mG$, we call the steady-state of $q_i$ an endemic state of $\mG_i$, and we denote it by $q_i^{\star}$. Hence, the endemic state emerging over the entire network is given by $p^\star = [q_1^{\star T},\hdots, q_N^{\star T}]^T$. 

We first state some results about the special case where the network topology is given by a DAG.

\begin{proposition} \label{lem::DAG1}
Let $\mG = (\mV,\mE)$ be a DAG and suppose $\delta_i>0$ for all $i \in \mV$. Then the disease-free equilibrium is the unique equilibrium. Moreover, this equilibrium is GAS.
\end{proposition}
\begin{pf}
Let us denote the steady-state of (\ref{eqn::SISmodelMatrix}) by $p(\infty)$. The steady-state equation for the source nodes of the DAG is of the form $0 = -\delta_ip_i(\infty)$, $i \in \mSo$, which implies that $p_i(\infty) =0$ for all source nodes. For a node $i \in \mSoT$, its steady-state equation can be written as $0 =  -\delta_ip_i(\infty) + (1-p_i(\infty)) \sum_{j \in \mSo} a_{ij}\beta_j p_j(\infty)$. The sum evaluates to zero, and again we obtain $p_i(\infty)=0$. By repeating this argument, we conclude that $p_i(\infty) = 0$, for all $i \in \mSoT$. By propagating this argument all the way to the sink nodes, we conclude that zero is the unique solution of the steady-state equation.

Next, we prove the second statement. In a DAG, the dynamics of the source nodes become  $\dot{p}_i = -\delta_i p_i$, $i \in \mSo$. Hence, all source nodes are globally exponentially stable. Let $v_i := \sum_{j\in \mSo} a_{ij}\beta_j p_j$, and define the following linear dynamical system for all $i \in \mSoT$
$$
\dot{\bar{p}}_i = -\delta_i \bar{p}_i + v_i, \quad \bar{p}_i(0) = p_i(0).
$$
Then, we have from (\ref{eqn::SISmodelMatrix}) that $\dot{p}_i \leq \dot{\bar{p}}_i$, for all $i \in \mSoT$. By the comparison lemma, it follows that $p_i\leq \bar{p}_i$, for all $t$ and all $i \in \mSoT$. It is well-known that if the input of an exponentially stable linear system converges to zero, its state converges to zero. Thus, since $v_i$ converges to zero, $\bar{p}_i$ must also converge to zero, for all $i \in \mSoT$. Since $p_i\geq 0$, we conclude that $p_i$ converges to zero for all $i \in \mSoT$. The proposition follows by repeating this argument for the remaining nodes in the graph.
 \qed
\end{pf}
We begin by studying the existence, uniqueness, and the stability properties of an endemic state over a weakly connected digraph consisting of two SCCs; the generalization to multiple SCCs is straightforward.
\begin{proposition} \label{prop::SCC}
Let $\mG_i=(\mV_i,\mE_i)$ be an SCC, $i \in [N]$, and let $q_i^\star$ be its endemic state equilibrium. If $q_{i,i_1}^\star > 0$ for some $i_1\in \mV_i$, then $q_i^\star \gg 0$.
\end{proposition}
\begin{pf}
Let $i_1 \in \mV_i$ be a node with $q_{i,i_1}^\star > 0$. Since $\mG_i$ is strongly connected, for any node $i_m \in \mV_i$, where $m$ is an integer satisfying $m \leq |\mV_i|$, there exists a directed path from node $i_1$ to node $i_m$. Let $i_2\in\mV_i$ be a node along this path such that $(i_1,i_2) \in \mE_i$. It follows from (\ref{eqn::pStar}), that $q_{i,i_2}^\star>0$. By the same argument, it follows that $q_{i,i_k}^\star >0$ for every node $i_k \in \mV_i$ along the directed path from $i_1$ to $i_m$, including $i_m$. Since nodes $i_1$ and $i_m$ were arbitrary, the proof is complete. \qed
\end{pf}

Let $\mR_0^i := \rho(D_i^{-1}A_i^TB_i)$ be the basic reproduction number corresponding to $\mG_i$. We have the following existence and uniqueness result.
\begin{theorem} \label{thm::existUniqueWeak}
Let $\mG=(\mV,\mE)$ be a weakly connected digraph consisting of two SCCs
$\mG_1$, $\mG_2$ such that $\mG_1 \prec \mG_2$.
Assume that $q_i(0) \neq 0$ for all $i\in [2]$. Then the following statements hold:
\begin{enumerate}[(i)]
\item If $\mR_0^1>1$, and $\mR_0^2$ being arbitrary, then $p = 0$ and $p^\star = [q_1^{\star T},q_2^{\star T}]^T$ are the only possible equilibrium points over $\mG$, where $q_1^\star$ and $q_2^\star$ are unique strong endemic equilibrium points over $\mG_1$ and $\mG_2$, respectively.

\item If $\mR_0^1\leq 1$ and $\mR_0^2>1$, then $p = 0$ and $p^\star = [0^T,q_2^{\star T}]^T$ are the only possible equilibrium points over $\mG$, where $q_2^\star$ is a unique strong endemic equilibrium point over $\mG_2$.

\item If $\mR_0^i\leq 1$, $i \in [2]$, then $p=0$ is the only possible equilibrium over $\mG$.
\end{enumerate}
\end{theorem}
\begin{pf}
In all the cases, the fact that $p=0$ is an equilibrium point follows directly from the structure of the dynamics. Since $\mG_1 \prec \mG_2$, we have $c_1 = 0$, i.e., the dynamics of the nodes in $\mG_1$ are not affected by those in $\mG_2$.

We first prove (i). First, consider the case when $\mR_0^2 > 1$. Since $\mR_0^1>1$ and $\mG_1$ is an SCC, we conclude by Theorems \ref{prop::pStarUniqueDirected} and \ref{thm::pStarGASDirected} that there exists a strong endemic state $q_1^\star \gg 0$ over $\mG_1$, which is GAS, assuming that $q_1(0) \neq 0$. Hence, $c_2$ converges to $c_2^\star :=  A_{12}^TB_2q^\star_1$, which is a nonnegative vector. We can now write the steady-state equation for $\mG_2$ as

\begin{equation} \label{eqn::steadyQ2}
(A_2^TB_2-D_2)q_2 - Q_2A_2^TB_2q_2 + (I-Q_2) c^\star_2 = 0,
\end{equation}
or
\[
A_2^TB_2 q_2- \diag(A_2^TB_2q_2)q_2 -(D_2+C^\star_2)q_2 +c_2^\star = 0,
\]
where $C^\star_2 = \diag(c^\star_2)$. Define $G_2= D_2+C_2$, and note that this is an invertible diagonal matrix because $D_2$ is a strictly positive diagonal matrix. We then conclude that
\[
G_2^{-1}A_2^TB_2 q_2- (I+\diag(G_2^{-1}A_2^TB_2q_2))q_2 +G_2^{-1}c_2^\star = 0,
\]
or
\begin{equation} \label{eqn::proof1G2}
q_2 = (I+\diag(G_2^{-1}A_2^TB_2q_2))^{-1}G_2^{-1}(A_2^TB_2 q_2+c_2^\star).
\end{equation}
Since $\mG_2$ is an SCC, $A_2$ is irreducible, and therefore $G_2^{-1}A_2^TB_2$ is irreducible as well. Furthermore, we have $G_2^{-1}c_2^\star \ll 1$ by construction. It then follows by Theorem \ref{thm::FP} in the Appendix that there exists a unique strong endemic state $q_2^\star$ over $\mG_2$. From (\ref{eqn::sspStarDirected}), it follows that the steady-state of any node in $\mG_2$ that is connected to a node in $\mG_1$ is strictly positive. Then, it follows from Proposition \ref{prop::SCC} that $[q_1^\star,0]$ cannot be an equilibrium over $\mG$, and $[q_1^{\star T},q_2^{\star T}]^T$ is the unique equilibrium over $\mG$ in this case.

When $\mR_0^2 \leq 1$, it follows from~\eqref{eqn::sspStarDirected} that the steady-state of any node in $\mG_2$ that is connected to a node in $\mG_1$ is strictly positive. Hence, by Proposition \ref{prop::SCC}, there exists a strong endemic state $q_2^\star$ over $\mG_2$. Finally, and because the steady-state equation over $\mG_2$ is given by (\ref{eqn::proof1G2}), it follows from Proposition \ref{prop::Tunique} in the Appendix that $q_2^\star$ must be unique.

For (ii), since $c_1 = 0$ and $\mR_0^1\leq 1$, it follows by Proposition \ref{prop::unstableDirected} and Theorem \ref{prop::pStarUniqueDirected} that the only valid equilibrium over $\mG_1$ is $q_1 = 0$, which is GAS. Hence, in steady-state, $\mG_2$ can be viewed as an isolated irreducible graph, and it follows from Theorems \ref{prop::pStarUniqueDirected} and \ref{thm::pStarGASDirected} that there exists a unique strictly positive equilibrium $q_2^\star$ over $\mG_2$.

Finally, for (iii), and similar to (ii), the only possible equilibrium over $\mG_1$ is $q_1=0$, which is GAS. This in turn leads to having $c_2^\star=0$, and since $\mR_0^2\leq 1$, the only possible equilibrium over $\mG_2$ is $q_2 = 0$.  \qed
\end{pf}

From (ii), we conclude that a weak endemic state could emerge over weakly connected graphs. A strong endemic state could emerge in case (i), and the disease-free state is the only possible equilibrium in case (iii). It is important to note that the endemic state $q_2^\star$ resulting in cases (i) and (ii) are not necessarily the same.

Next, we study the stability properties of weak and strong endemic equilibria.

\begin{theorem}\label{thm::ISSRo<1}%\longthmtitle{global asymptotic stability of cascaded SCCs}
Let $\mG=(\mV,\mE)$ be a weakly connected digraph consisting of two
SCCs $\mG_1$, $\mG_2$ such that $\mG_1 \prec \mG_2$.
Assume that $q_i(0) \neq 0$ for all $i \in [2]$. Then, $\mG_2$ is input-to-state stable (ISS). {Further, for all possible values of $\mR_0^1$ and $\mR_0^2$, the resulting equilibrium over $\mG$ is GAS.}
\end{theorem}
\begin{pf}
First, note that the dynamics over $\mG_1$ are not affected by $\mG_2$. Hence, the global asymptotic stability of the equilibrium (disease-free or strong endemic, depending on the value of $\mR_0^1$) over $\mG_1$ follows immediately. We will start by proving that $\mG_2$ is ISS for different values of $\mR_0^1$ and $\mR_0^2$. Consider the following cases.

\emph{(i) $\mR_0^2<1$:} In this case, we have $\mu(A_2^TB_2-D_2)<0$, and therefore the matrix $A_2^TB_2-D_2$ is Hurwitz. Since it is also Metzler, it follows from Proposition \ref{prop::rantzer} that there exists a positive diagonal matrix $R$ which satisfies
$$
(A_2^TB_2-D_2)^TR +R(A_2^TB_2-D_2)  = -K,
$$
where $K$ is a positive definite matrix. Similar to the proof of Proposition \ref{prop::unstableDirected}, consider the Lyapunov function $V_R(q_2) = q_2^TRq_2$. We have
\begin{eqnarray*}
\Lie_{\tilde{\Phi}_2}V_R(q_2) & = & q_2^T((A_2^TB_2-D_2)^TR+R(A_2^TB_2-D_2))q_2 \\
& -& 2q_2^TRQ_2A_2^TB_2q_2 + 2q_2^TR(I-Q_2)c_2\\
&\leq&  -q_2^TKq_2 + 2q_2^TRc_2,
\end{eqnarray*}
where the inequality follows because $q_2^TRQ_2A_2^TB_2q_2 \geq 0$, for all $q_2 \in [0,1]^n$, and $q_2^TRQ_2c_2 \geq 0$, for all $c_2,q_2 \in [0,1]^n$. Let $0< \epsilon < 1$. We can then write
\begin{eqnarray*}
\Lie_{\tilde{\Phi}_2}V_R(q_2) & \leq & -(1-\epsilon)q_2^TKq_2 - \epsilon q_2^TKq_2 + 2q_2^TRc_2.
\end{eqnarray*}
We will prove that there exists a class $\mK_\infty$ function,
$\chi$, such that $- \epsilon q_2^TKq_2 + 2q_2^TRc_2 \leq 0$ for $\|q_2\|_2 \geq \chi(\|c_2\|_2)$. To this end, note that $q_2^TRc_2 \leq \|R\|_2 \cdot \|q_2\|_2\cdot \|c_2\|_2$. Also, because $K$ is positive definite, we can write $q_2^TKq_2\geq \lambda_n(K)\|q\|_2^2>0$. Define
$\chi(r) := \frac{2\|R\|_2\cdot r}{\epsilon \lambda_n(K)} $, where $ r \in \mathbb{R}$. We then have $- \epsilon q_2^TKq_2 + 2q_2^TRc_2 \leq 0$ for $\|q_2\|_2\geq \chi(\|c_2\|_2)$, and hence
\begin{eqnarray*}
\Lie_{\tilde{\Phi}_2}V_R(q_2) & \leq  -(1-\epsilon)q_2^TKq_2, \quad \|q_2\|_2 \geq \chi(\|c_2\|_2).
\end{eqnarray*}
This implies that the system $\mG_2$ is ISS when $\mR_0^2 < 1$ and $\mR_0^1$ is arbitrary. 

\emph{(ii) $\mR_0^2=1$:} Following the same reasoning in the proof of Proposition \ref{prop::unstableDirected}, we conclude that there exists a positive diagonal matrix $S$ such that $(A_2^TB_2-D_2)^TS+S(A_2^TB_2-D_2)$ is negative semidefinite. Then, using the Lyapunov function $V_S(q_2)=q_2^TSq_2$, we can write
\begin{eqnarray*}
\Lie_{\tilde{\Phi}_2}V_S(q_2) & \leq & -2q_2^TQ_2SA_2^TB_2q_2 + 2q_2^TSc_2 \\
& \leq & -q_2^TQ_2SA_2^TB_2q_2 + 2\sqrt{n} \|S\|_2\cdot \|c_2\|_2,
\end{eqnarray*}
where the second inequality follows by using the bound $\|q_2\|_2 \leq \sqrt{|\mV_2|} \leq \sqrt{n}$. Define the function $\rho:\real \to \real$ as $\rho(\|c_2\|_2) = 2\sqrt{n} \|S\|_2\cdot \|c_2\|_2$, and note that $\rho \in \mK_\infty$ since it is linear in $\|c\|_2$. Define the function $g:\real_{\geq 0}^n\to \real$ as $g(q_2) = 2q_2^TQ_2SA_2^TB_2q_2$. Following similar steps to those in the proof of Proposition \ref{prop::unstableDirected}, we can show that $g(q_2) = 0$ if and only if $q_2 = 0$. Note that $g(q_2)>0$ for all $q_2 \in \real^n_{\geq 0}$ such that $q_2 \neq 0$. Furthermore, the function $g$ is continuous and radially unbounded. Hence, it follows by \cite[Lemma 4.3]{khalil2002nonlinear} that there exists a class $\mK_\infty$ function $\alpha:\real \to \real$ such that $g(q_2) \geq \alpha(\|q_2\|_2)$. We therefore have
\[
\Lie_{\tilde{\Phi}_2}V_S(q_2) \leq -\alpha(\|q_2\|_2) + \rho(\|c_2\|_2).
\]
As a result, it follows from \cite[Remark 2.4]{sontag1995characterizations} that the system $\mG_2$ is ISS when $\mR_0^2 = 1$ and $\mR_0^1$ is arbitrary.

\emph{(iii) $\mR_0^2 > 1$}: Define the state $\tilde{q}_2 = q_2 - q_2^\star$, and the control input $\tilde{c}_2 = c_2 - c_2^\star$, where $c_2^\star$ was defined in the proof of Theorem \ref{thm::existUniqueWeak} as the steady-state of $c_2$. Let $\tilde{Q}_2=\diag(\tilde{q}_2)$, $Q_2^\star = \diag(q_2^\star)$, and $C_2^\star = \diag(c_2^\star)$. The dynamics of $\tilde{q}_2$ can then be written as
\begin{eqnarray}
\dot{\tilde{q}}_2 & = & (A_2^TB_2-D_2)(\tilde{q}_2+q_2^\star) -(\tilde{Q}_2+Q_2^\star)A_2^TB_2(\tilde{q}_2 +q_2^\star) \nonumber \\
&&+(I-\tilde{Q}_2-Q^\star_2)(\tilde{c}_2+c_2^\star) \nonumber \\
& = &(-D_2+(I-Q_2^\star)A_2^TB_2)\tilde{q}_2-\tilde{Q}_2A_2^TB_2q_2 \nonumber \\
&&+(I-Q_2)\tilde{c}_2 -\tilde{Q}_2c_2^\star \label{eqn::auxQ2tilde1}\\
& = &(-D_2-C_2^\star+(I-Q_2^\star)A_2^TB_2)\tilde{q}_2-\tilde{Q}_2A_2^TB_2q_2 \nonumber \\
&&+(I-Q)\tilde{c}_2, \label{eqn::auxQ2tilde2}
\end{eqnarray}
where \eqref{eqn::auxQ2tilde1} follows from the steady-state equation (\ref{eqn::steadyQ2}) evaluated at $q_2=q_2^\star$, and \eqref{eqn::auxQ2tilde2} follows because $\tilde{Q}_2c^\star_2 = C^\star_2\tilde{q}_2$.

Next, define the matrix $\tilde{\Lambda}(q^\star_2) = -D_2-C_2^\star+(I-Q_2^\star)A_2^TB_2$, which is Metzler since its off-diagonal entries are nonnegative. Since $\mG_2$ is an SCC, the matrix $\tilde{\Lambda}(q^\star_2)$ is also irreducible. We wish to study the sign of $\mu\left(\tilde{\Lambda}(q^\star_2)\right)$. Using the steady-state equation (\ref{eqn::steadyQ2}) evaluated at $q_2=q_2^\star$, it follows that $\tilde{\Lambda}(q^\star_2) q_2^\star = -c_2^\star$, where we recall that $c_2^\star \succeq 0$. Consider the following two cases.

\emph{(iii.a) $\mR_0^1\leq 1$ and $\mR_0^2 > 1$}: In this case, the disease-free state is GAS over $\mG_1$; see Proposition \ref{prop::unstableDirected}. Then, $c_2^\star = 0$, and $\tilde{\Lambda}(q^\star_2) q_2^\star = 0$. Since $q_2^\star$ is strictly positive, it follows from Theorem \ref{thm::PFMI} that $\mu\left(\tilde{\Lambda}(q^\star_2)\right) = 0$. Thus, it follows from Lemma \ref{lem::semi} that there exists a positive diagonal matrix $R$ such that the matrix $\tilde{\Lambda}(q^\star_2)^TR+ R\tilde{\Lambda}(q^\star_2)$ is negative semidefinite. Consider the Lyapunov function $V_R(\tilde{p})= \tilde{p}^TR\tilde{p}$. We have
\begin{eqnarray}
\Lie_{\tilde{\Phi}_2} V_R(\tilde{p}) & = & \tilde{q}_2^T(\tilde{\Lambda}(q_2^\star)^TR + R\tilde{\Lambda}(q_2^\star))\tilde{q}_2 - 2\tilde{q}_2^T\tilde{Q}_2RA_2^TB_2q_2 \nonumber \\
&& + 2\tilde{q}_2^TR(I-Q_2)\tilde{c}_2  \nonumber  \\
& \leq & - 2\tilde{q}_2^T\tilde{Q}_2RA_2^TB_2q_2 + 2\tilde{q}_2^TR(I-Q_2)\tilde{c}_2 \nonumber  \\
& \leq & - 2\tilde{q}_2^T\tilde{Q}_2RA_2^TB_2q_2 + 4\sqrt{n} \|R\|_2\cdot \|\tilde{c}_2\|_2, \label{eqn::auxUpperBound}
\end{eqnarray}
where the last inequality follows from $\|\tilde{q}_2\|_2 \leq \|q_2\|_2 + \|q_2^\star\|_2 \leq 2\sqrt{n}$, and the fact that $\|I-Q_2\|_2 \leq 1$. Define the scalar function $\rho(\|\tilde{c}_2\|_2) := 4\sqrt{n} \|R\|_2\cdot \|\tilde{c}_2\|_2$, and note that $\rho \in \mK_\infty$, since it is linear in $\|\tilde{c}_2\|_2$. Following similar steps to those in the proof of Theorem \ref{thm::pStarGASDirected}, one can show that $\tilde{q}_2^T\tilde{Q}_2RA_2^TB_2q_2 =0$ if and only if $\tilde{q}_2 = 0$. Then, using the same reasoning as in the proof of Theorem \ref{thm::ISSRo<1}, we conclude that there exists a class $\mK_\infty$ function $\alpha:\mathbb{R}\to \mathbb{R}$ such that $2\tilde{q}_2^T\tilde{Q}_2RA_2^TB_2q_2 \geq \alpha(\|\tilde{q}_2\|_2)$. We therefore have $\Lie_{\tilde{\Phi}_2} V_R(\tilde{p}) \leq - \alpha(\|\tilde{q}_2\|_2) + \rho(\|\tilde{c}_2\|_2)$, and it follows from \cite[Remark 2.4]{sontag1995characterizations} that the system $\mG_2$ is input-to-state-stable when $\mR_0^1 \leq 1$ and $\mR_0^2 > 1$.

\emph{(iii.b) $\mR_0^1> 1$ and $\mR_0^2>1$}: In this case, the endemic state is GAS over $\mG_1$; see Theorem \ref{thm::pStarGASDirected}. Then, $c_2^\star \succ 0$, and $\tilde{\Lambda}(q^\star_2) q_2^\star \prec 0$. Since $q_2^\star$ is strictly positive, it follows from \cite[Theorem 2.4]{fall2007epidemiological} that $\mu\left(\tilde{\Lambda}(q^\star_2)\right) < 0$; therefore, $\tilde{\Lambda}(q^\star_2)$ is Hurwitz. Thus, it follows from Proposition \ref{prop::rantzer}(iv) that there exists a positive diagonal matrix $S$ such that the matrix $\tilde{\Lambda}(q^\star_2)^TS+ S\tilde{\Lambda}(q^\star_2)$ is negative definite. Hence, using $V_S(\tilde{p})= \tilde{p}^TS\tilde{p}$, one can derive the same bound as in \eqref{eqn::auxUpperBound}, with $R$ replaced with $S$, and by repeating the same steps as above, one can show that $\mG_2$ is input to state stable when $\mR_0^1 > 1$ and $\mR_0^2 > 1$.

Since $\mG_1$ is GAS, and $\mG_2$ is ISS, it follows from \cite[Lemma 4.7]{khalil2002nonlinear} that the equilibrium of the cascaded system is GAS. In particular, when $\mR_0^2 \leq 1$ and $\mR_0^1 \leq 1$, it follows from Theorem \ref{thm::existUniqueWeak}(iii) that the disease-free state is GAS. When $\mR_0^2 \leq 1$ and $\mR_0^1 > 1$, it follows from Theorem \ref{thm::existUniqueWeak}(i) that the strong endemic equilibrium $[q_1^{\star T},q_2^{\star T}]^T$ is GAS, assuming that $q_i(0) \neq 0$ for all $i \in [2]$. When $\mR_0^2 > 1$ and $\mR_0^1 \leq 1$, it follows from Theorem \ref{thm::existUniqueWeak}(ii) that the weak endemic state $[0^T,q_2^{\star T}]^T$ is GAS, assuming that $q_2(0) \neq 0$. Finally, when when $\mR_0^2 > 1$ and $\mR_0^1 > 1$, it follows from Theorem \ref{thm::existUniqueWeak}(i) that the strong endemic state $[q_1^{\star T},q_2^{\star T}]^T$ is GAS, assuming that $q_i(0)\neq 0$ for $i \in [2]$. \qed
\end{pf}

The following corollary is an immediate consequence of Theorems \ref{thm::existUniqueWeak} and \ref{thm::ISSRo<1}.
\begin{corollary}
Let $\mG=(\mV,\mE)$ be a weakly connected digraph consisting of $N$ SCCs ordered as $\mG_1 \prec \hdots \prec \mG_N$. Assume that $q_i(0) \neq 0$ for all $i \in [n]$.
\begin{enumerate}[(i)]
\item If $\mR_0^i\leq 1$ for all $i \in [N]$, then the disease-free state is GAS.

\item If $\mR_0^k> 1$ for some $k \in [N]$, and $\mR_0^i \leq 1$ for $i \in \{1,\hdots,k-1 \}$, then the endemic state $p^\star = [0,\hdots,0,q_k^{\star T},\hdots,q_N^{\star T}]^T$ is GAS.
%\item If $\mR_0^k> 1$ for some all $i \in [N]$, then the strong endemic state $p^\star = [q_1^\star,\hdots,q_N^\star]$ is GAS.
\end{enumerate}
\end{corollary}

\section{Numerical Studies} \label{sec::simulations}
\begin{figure}[htp]
\vspace{2mm}
\centering
\includegraphics[width=1\linewidth]{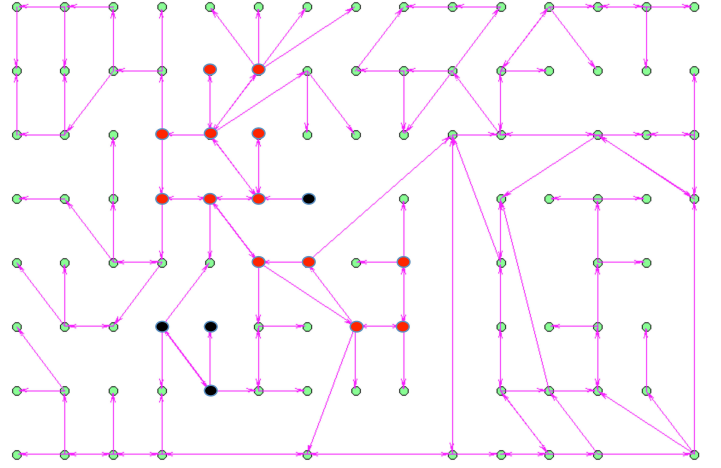}
\caption{The Pajek GD99c network. The ``red" nodes belong to $\mG_1$ for which $\mR_0^1 >1$. The ``black" nodes are the only ones with no direct path from $\mG_1$.}
\label{fig::DAG}
\vspace{-2mm}
\end{figure}

%For digraphs, 
We demonstrate the emergence of a weak endemic state over the Pajek GD99c network \cite{pajekGD99c}, which is a weakly connected directed network shown in Fig. \ref{fig::DAG}. The network consists of $105$ nodes and it contains $66$ SCCs. The nodes marked ``red" in Fig. \ref{fig::DAG} constitute an SCC, which we refer to as $\mG_1$. We will select the curing rates over $\mG_1$ to be low in order to make $\mR_0^1 > 1$. For the remaining nodes, we will set $\delta_i = \sum_{j\neq i}a_{ji}\beta_j +0.5$, which is a sufficient condition to ensure $\mR_0^i < 1$ \cite{KhanaferBasarGharesifardACC14}. The infection rates $\beta_i$ and the weights $a_{ij}$ are all selected to be equal to $1$. There are only $4$ nodes for which there is no directed path from $\mG_1$, and they are marked ``black" in Fig. \ref{fig::DAG}. The initial infection profile is selected at random.

\begin{figure}[htp]
\centering
\includegraphics[width= .90\linewidth]{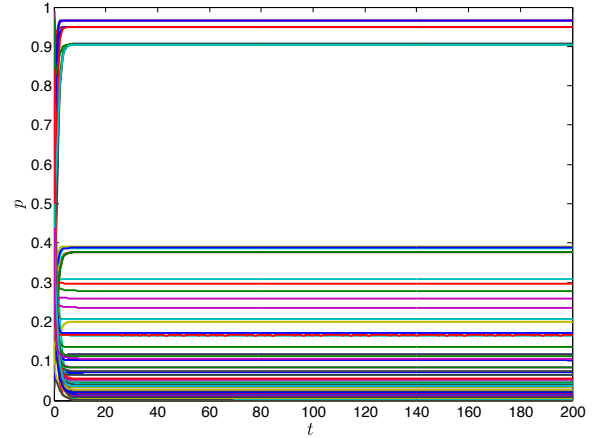}
\caption{Infection probabilities of the nodes.}
\label{fig::bigNet}
\end{figure}

Fig. \ref{fig::bigNet} plots the state trajectories. By examining the histogram of the values to which the state converges, we notice that there are $13$ nodes with high infection probabilities, and those are the nodes comprising $\mG_1$. Note that $\mG_1$ is asymptotically stable even though it takes input from other SCCs, as shown in the figure, and $\mR_0^1>1$. There are $4$ nodes that become healthy, and those are the ``black" nodes which are not reached by a directed path from $\mG_1$. The remaining nodes all have positive infection probabilities with varying levels depending on their distance from $\mG_1$, with the nodes that are farthest from $\mG_1$ enjoying the lowest infection probabilities.

\begin{figure}[htp]
\vspace{2mm}
\centering
\includegraphics[width= .90\linewidth]{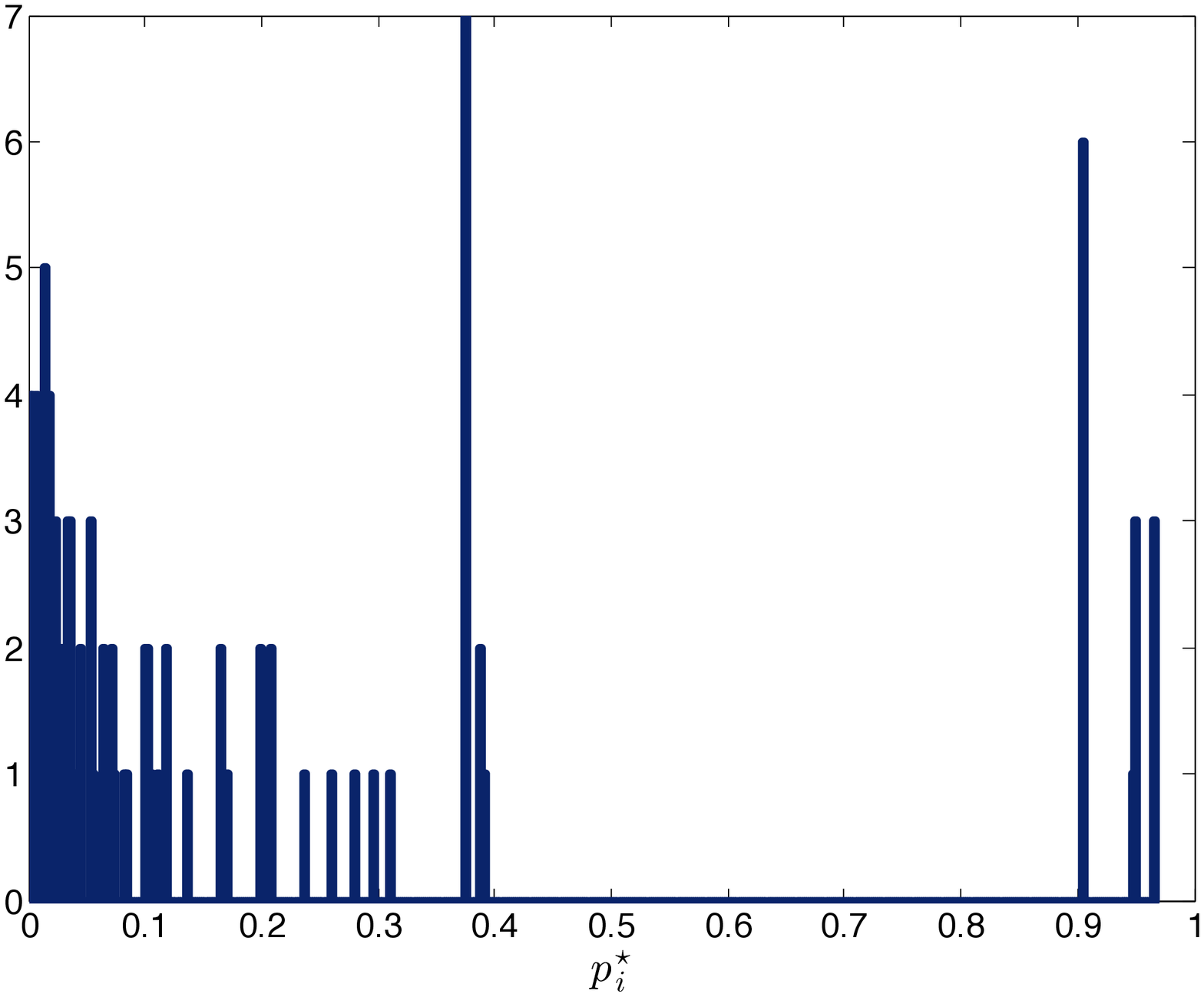}
\caption{A histogram of the endemic state value across the network.}
\label{fig::histBigNet}
\vspace{-2mm}
\end{figure}

Next, we will demonstrate the global asymptotic stability of $p^\star$ over connected undirected graphs, which follows from Theorem \ref{thm::pStarGASDirected}. The infection rates, the edge weights, and the initial infection profile were generated randomly. The curing rates were selected such that $\mR_0 > 1$.

\begin{figure}[htp]
\centering
\includegraphics[width= 8 cm]{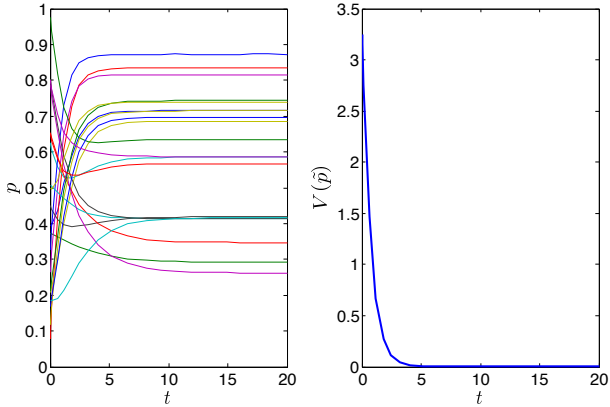}
\caption{Stabilization of a ring graph with $20$ nodes.}
\label{fig::10ring}
\end{figure}

Fig. \ref{fig::10ring} shows the state of a ring graph with $20$ nodes. The figure also plots the Lyapunov function $V(\tilde{p}) = \frac{1}{2}\tilde{p}^T\tilde{p}$. As claimed, the system converges to the strictly positive state $p^\star$, and the Lyapunov function decays monotonically to zero.

\begin{figure}[htp]
\centering
\includegraphics[width= 8 cm]{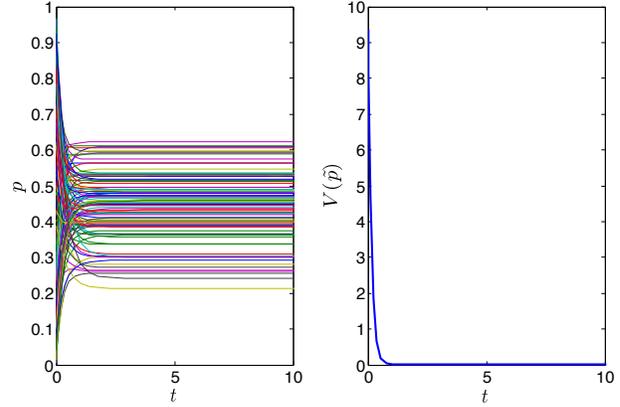}
\caption{Stabilization of a random graph with $100$ nodes and $1704$ edges.}
\label{fig::100rg}
\end{figure}

Fig. \ref{fig::100rg} shows the same simulation for a connected undirected random graph with $100$ nodes. The probability that an edge occurs in the graph was selected to be $\tfrac{3}{10}$. The specific graph realization used in this experiment contained $1704$ edges. Again, we observe that the state converges to $p^\star$. It is interesting to note that convergence here is faster than the case of the ring graph.

\section{Conclusion} \label{sec::conclusion}
We have utilized tools from positive systems theory to establish the stability properties of the $n$-intertwined Markov model over digraphs. For strongly connected digraphs, we have proved that when the basic reproduction number is less than or equal to $1$, the disease-free state is GAS. When the basic reproduction number is greater than $1$, however, we have shown that the endemic state is GAS, and that locally around this equilibrium, the convergence is exponentially fast. Furthermore, we have studied the stability properties of weakly connected graphs. By viewing an arbitrary weakly connected graph as a cascade of SCCs, we were able to establish the existence and uniqueness of weak and strong endemic states. We have also studied the stability properties of weakly connected graphs using input-to-state stability. Finally, we have proposed a dynamical model that describes the interaction among nodes in an infected network as a concave game and demonstrated that the $n$-intertwined Markov model is a special case of our model. This alternative description provides a new condition, which can be checked collectively by agents, for the stability of the disease-free equilibrium.

Future work will focus on studying the stability properties of the SIS dynamics over time-varying networks and designing optimal dynamic curing mechanisms. 
%\margin{Reference list needs to be updated: 1) Reference~[1] does not have an author. 2) There is a
%general inconsistency with referring to first name or not (the practice of journals is not to use first names,~[3] is one that you have done correctly.) 3) I think you need to adopt a system for labeling references. Here is one that is common within control people I have worked with Reference~[3] is labeled as TB-GIO:99 and~[24] as HRT:11. Sometime you can use extra things to denote better where the paper belongs. For example,~[16] can be KSN-RS:10-tac. Please change them all to make it systematic for future.}

\bibliographystyle{abbrv}        % Include this if you use bibtex
\bibliography{references}           % and a bib file to produce the

\appendix

\newcounter{mycounter}
\renewcommand{\themycounter}{A.\arabic{mycounter}}
\newtheorem{propositionappendix}[mycounter]{Proposition}
\newtheorem{theoremappendix}[mycounter]{Theorem}
\newtheorem{lemmaappendix}[mycounter]{Lemma}

\section{Appendix}\label{sec:appendix}
In this Appendix, we collect and prove some results pertinent to the development in the main body of the paper. We start with the next result which is key in proving some of the results in Sections \ref{sec::directed} and \ref{sec::weakly}. 

\begin{lemmaappendix} \label{lem::semi}
Let $X\in \mathbb{R}^{n\times n}$ be an irreducible Metzler matrix such that $\mu(X) = 0$. Then, there exists a positive diagonal matrix $R\in \mathbb{R}^{n\times n}$ such that the matrix $X^TR+RX$ is negative semidefinite.
\end{lemmaappendix}
\begin{pf}
From Theorem \ref{thm::PFMI}, it follows that there exists a vector $\nu \in \mathbb{R}^{n \times n}$ such that $\nu \gg 0 $ and $X \nu = 0$. Since $\sigma(X) = \sigma(X^T)$, we have $\mu(A^T) = 0$. Using Theorem \ref{thm::PFMI} again, we conclude that there exists a vector $\xi \in \mathbb{R}^{n\times n}$ such that $\xi \gg 0$ and $X^T\xi = 0$. Let $R \in \mathbb{R}^{n \times n}$ be a positive diagonal matrix defined with $R_{ii} = \xi_i/\nu_i$, for all \iset. Consider now the matrix $X^TR+RX$. The matrix $ RX $ is Metzler, since $R$ is a positive diagonal matrix. For the same reason, and because $X$ is irreducible, we conclude that $RX$ is irreducible. By a similar argument, $X^TR$ is also an irreducible Metzler matrix. Since the sum of two Metzler matrices is Metzler, the matrix $X^TR + RX$ is Metzler. Also, because both $RX$ and $X^TR$ are Metzler and irreducible, the matrix $X^TR + RX$ is also irreducible. Further, by construction, we have $(X^TR + RX)\nu =  X^TR\nu = X^T\xi= 0$. Since $X^TR + RX$ is symmetric, it has real eigenvalues, and since $\nu$ is strictly positive, it follows from Theorem \ref{thm::PFMI} that $X^TR + RX$ is negative semidefinite. \qed
\end{pf}

Next, we prove an instrumental result, which can be thought of as a non-homogeneous extension of a result of~\cite{fall2007epidemiological}. We start by providing two key properties of the continuous mapping $T: [0,1]^n \to [0,1]^n$ defined as
\begin{equation}
T(p) := (I+\diag(Xp))^{-1}(Xp+y).
\end{equation}

\begin{propositionappendix} \label{prop::Tmonotone}
Let $X \in \mathbb{R}^{n \times n}$ be a nonnegative matrix, and let $y \in \mathbb{R}^n$ be a vector satisfying $0 \preceq y\ll \mathbf{1}$. Then, the mapping $T$ is monotonic.
\end{propositionappendix}
\begin{pf}
Let the vectors $p,q\in \mathbb{R}^n$ be such that $p \preceq q$. For $i \in [n]$, we have
\begin{eqnarray*}
T_i(p) & = & \frac{(Xp)_i + y_i}{1 + (Xp)_i} = 1 - \frac{1-y_i}{1 + (Xp)_i}  \\
& \leq & 1 - \frac{1-y_i}{1 + (Xq)_i} = T_i(q),
\end{eqnarray*}
where the inequality follows because $X$ is nonnegative. This implies that the mapping $T$ is monotonic. \qed
\end{pf}

\begin{propositionappendix} \label{prop::Tunique}
Let $X \in \mathbb{R}^{n \times n}$ be a nonnegative matrix, and let $y \in \mathbb{R}^n$ be a vector satisfying $0 \preceq y\ll \mathbf{1}$. If the mapping $T$ has strictly positive fixed point, then it must be unique.
\end{propositionappendix}
\begin{pf}
We will prove the claim by contradiction. Assume that there are two fixed points $p^\star, q^\star \in \mathbb{R}^n$, $p^\star \neq q^\star$.  We will first show that $p^\star \preceq q^\star$. To this end, define
\[
\eta := \max\limits_{i \in [n]} \frac{p^\star_i}{q^\star_i}, \quad k := \argmax\limits_{i \in [n]} \frac{p^\star_i}{q^\star_i}.
\]
Note that $p^\star \preceq \eta q^\star$. For $p^\star \preceq q^\star$ to hold, we must have $\eta \leq 1$; assume that, to the contrary, $\eta > 1$. Then, using Proposition \ref{prop::Tmonotone}, we have
\begin{eqnarray*}
p^\star_k & = & T_k(p^\star) \leq T_k(\eta q^\star) = \frac{\eta (Xq^\star)_k + y_k}{1+\eta (Xq^\star)_k} \\
& < &  \eta \frac{(Xq^\star)_k + y_k}{1+(Xq^\star)_k} = \eta T_k(q^\star) = \eta q^\star,
\end{eqnarray*}
where the strict inequality follows from the assumption that $\eta > 1$, and the last equality follows because $q^\star$ is a fixed point. By definition, we have $p^\star_k = \eta q^\star_k$. Hence, if $\eta>1$ were true, we would have $p_k^\star < \eta q_k^\star = p_k^\star$, which is a contradiction. Hence, we must have $\eta \leq 1$ and $p^\star \preceq q^\star$. By switching the roles of $p^\star$ and $q^\star$, and repeating the above steps with $\hat{\eta} = \max_{i \in [n]} \frac{q^\star_i}{p^\star_i}$ instead of $\eta$, we conclude that $p^\star \succeq q^\star$. Thus, $p^\star = q^\star$, and the fixed point is unique. \qed
\end{pf}

We are now ready to prove the main result.
\begin{theoremappendix} \label{thm::FP}
Let $X \in \mathbb{R}^{n \times n}$ be a nonnegative irreducible matrix such that $\rho(X) > 1$, and let $y \in \mathbb{R}^n$ be a vector satisfying $0 \preceq y\ll \mathbf{1}$. Then, the mapping $T: [0,1]^n \to [0,1]^n$ has a unique fixed point, which is strictly positive.
\end{theoremappendix}
\begin{pf}
We will prove that there exists a closed sub-interval of $(0,1)^n$ which is invariant under $ T $. By Theorem \ref{thm::PFNI}, it follows that $X$ has an eigenvector $v\gg 0$ satisfying $Xv = \rho(X)v$. Without loss of generality, we assume that $v\preceq 1$, which can be achieved by an appropriate scaling of the eigenvector corresponding to $\rho(X)$.

Define $\overline{\kappa} := \sqrt{\frac{\rho(X)+\subscr{y}{max}}{1+\rho(X)}}$, and note that $\overline{\kappa} < 1$. Let us choose $\overline{\epsilon}>0$ such that $ \overline{\kappa} \leq \overline{\epsilon} \subscr{v}{min}$. Note that with such a choice of $\overline{\epsilon}$, we can guarantee, for all $i\in [n]$, that $\overline{\epsilon}v_i < 1$, since $v_i \leq 1$ and $\overline{\kappa}  <1$. This choice of $\overline{\epsilon}$ implies that $\overline{\epsilon} v_i \geq \overline{\kappa}$ or $(\overline{\epsilon} v_i)^2 \geq \frac{\rho(X)+\subscr{y}{max}}{1+\rho(X)}$, for all $i\in [n]$. This in turn implies, for $i \in [n]$,
\begin{eqnarray}
\overline{\epsilon} v_i \geq \frac{1}{\overline{\epsilon} v_i} \cdot \frac{\rho(X)+y_i}{1+\rho(X)}  > \frac{\overline{\epsilon} \rho(X)v_i+y_i}{1+\overline{\epsilon} v_i\rho(X)}  = T_i(\overline{\epsilon} v_i),\label{eqn::Teps2}
\end{eqnarray}
where the last inequality follows since $\overline{\epsilon} v_i < 1$. We therefore have $T(\overline{\epsilon} v) < \overline{\epsilon} v$.

Define $\underline{\kappa} := \frac{\rho(X)+\subscr{y}{min}-1}{1+\rho(X)}$, and note that $\underline{\kappa} < 1$, as $\subscr{y}{min}<1$. Let us choose $\underline{\epsilon} > 0$ such that $0<\underline{\epsilon}\subscr{v}{max}  \leq \underline{\kappa}$. Then, for all $i \in [n]$, we have
\begin{equation*}
\underline{\epsilon} v_i\leq \frac{\rho(X)+y_i-1}{\rho(X)+1} <  \frac{\rho(X)+y_i-1}{\rho(X)}.
\end{equation*}
We thus have $\underline{\epsilon} \rho(X) v_i +1< \rho(X)+y_i$, for all $i\in [n]$. Equivalently, for all $i\in [n]$, we can write
\begin{eqnarray}
\underline{\epsilon} v_i < \underline{\epsilon} v_i \frac{\rho(X)+y_i}{\underline{\epsilon}\rho(X)v_i+1}  <  \frac{\underline{\epsilon} \rho(X)v_i+y_i}{\underline{\epsilon}\rho(X)v_i+1} = T_i(\underline{\epsilon} v), \label{eqn::Teps1}
\end{eqnarray}
where the second strict inequality holds since $\underline{\epsilon}v_i < \kappa<1$. We therefore have $T(\underline{\epsilon}v) > \underline{\epsilon} v$.

Since $v \gg 0$ and $\underline{\epsilon} >0$, we have $\underline{\epsilon} v \gg 0$. We also have that $\overline{\epsilon} > \underline{\epsilon}$ because
\begin{eqnarray*}
\overline{\epsilon} & \geq & \frac{\overline{\kappa}}{\subscr{v}{min}} > \frac{\overline{\kappa}^2}{\subscr{v}{min}} =  \frac{\rho(X)+\subscr{y}{max}}{\subscr{v}{min}(1+\rho(X))} \\
& >&  \frac{\rho(X)+\subscr{y}{min}-1}{\subscr{v}{max}(1+\rho(X))} = \frac{\underline{\kappa}}{\subscr{v}{max}} \geq \underline{\epsilon},
\end{eqnarray*}
where the first strict inequality follows because $\overline{\kappa} < 1$. This implies that $\underline{\epsilon} v \ll \overline{\epsilon} v$. Further, by construction, we have $\overline{\epsilon}v_i < 1$, for all $i\in [n]$, and therefore $ \overline{\epsilon} v \ll 1$. To summarize,  we have the following bounds: $0 \ll \underline{\epsilon} v \ll \overline{\epsilon} v \ll 1$.

We can now define the closed and bounded set
\[
K := \{p \in [0,1]^n: \epsilon_1v \preceq p \preceq \epsilon_2 v \} \subset (0,1)^n.
\]
By (\ref{eqn::Teps2}) and (\ref{eqn::Teps1}), and since $T$ is monotonic as proved in Proposition \ref{prop::Tmonotone}, we conclude that $T: K \to K$. Since $T$ is continuous, it follows from Brouwer's fixed-point theorem that there exists a strictly positive fixed point $p^\star \in K$ such that $T(p^\star) = p^\star$. Finally, it follows from Proposition \ref{prop::Tunique} that $p^\star$ must be unique. \qed
\end{pf}

\end{document}